\documentclass[
  ,final            
  ]
  {aipproc}

\layoutstyle{6x9}

\begin{document}

\title{ A relativistic model for neutrino pion production from nuclei
  in the resonance region }


\author{C.~Praet}{
  address={Proeftuinstraat 86, B-9000 Gent, Belgium}
}

\author{O.~Lalakulich}{
}

\author{N.~Jachowicz}{
}

\author{J.~Ryckebusch}{
}

\begin{abstract}

We present a relativistic model for electroweak pion production from
nuclei, focusing on the $\Delta$ and the second resonance region.
Bound states are derived in the Hartree approximation to the
$\sigma-\omega$ Walecka model.  Final-state interactions of the
outgoing pion and nucleon are described in a factorized way by means
of a relativistic extension of the Glauber model.  Our formalism
allows a detailed study of neutrino pion production through $Q^2$,
$W$, energy, angle and out-of-plane distributions.  

\end{abstract}

\maketitle

\begin{figure}
\label{Preliminary}
\includegraphics[width=.5\textwidth]{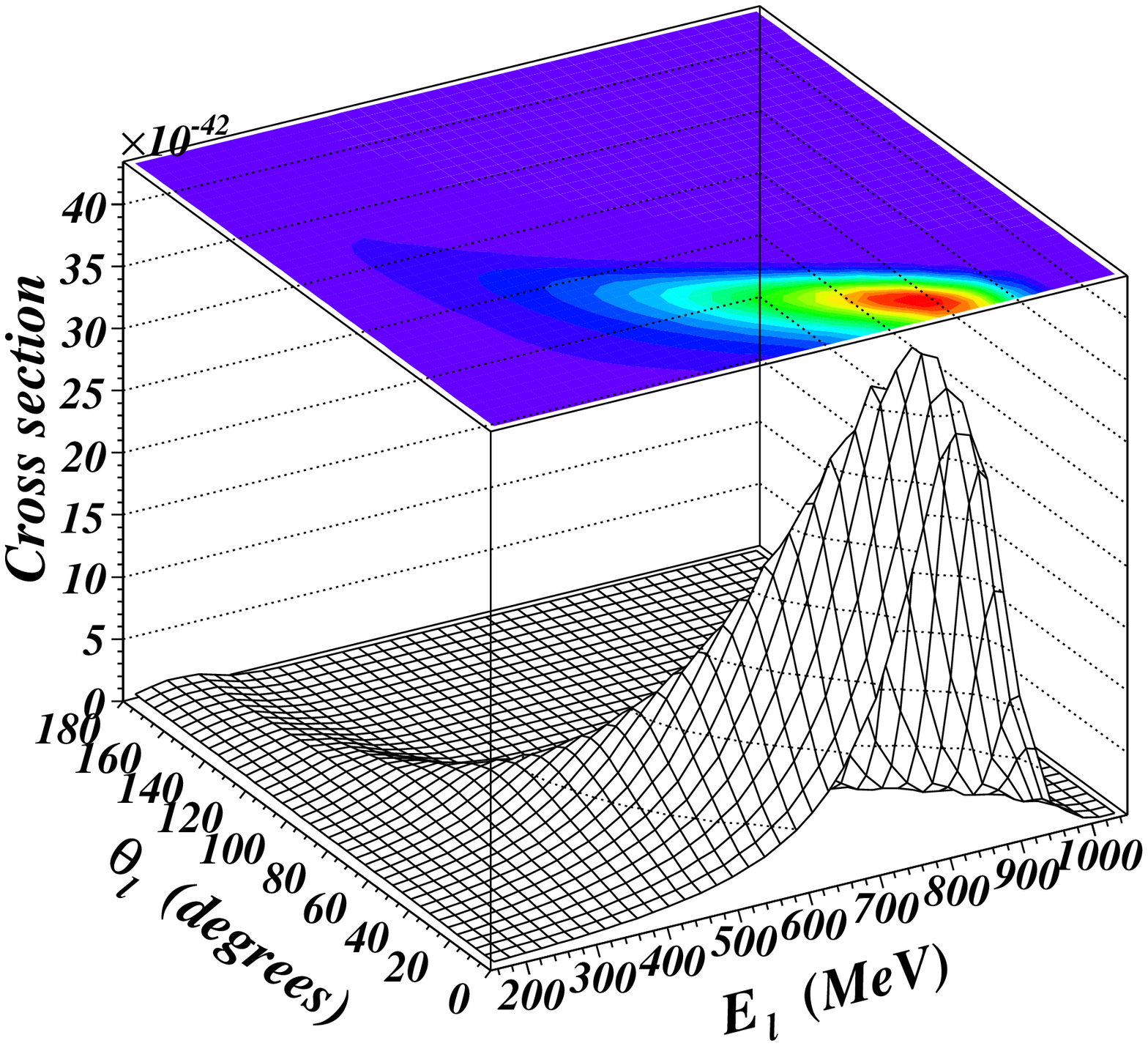}\hfill
\includegraphics[width=.5\textwidth]{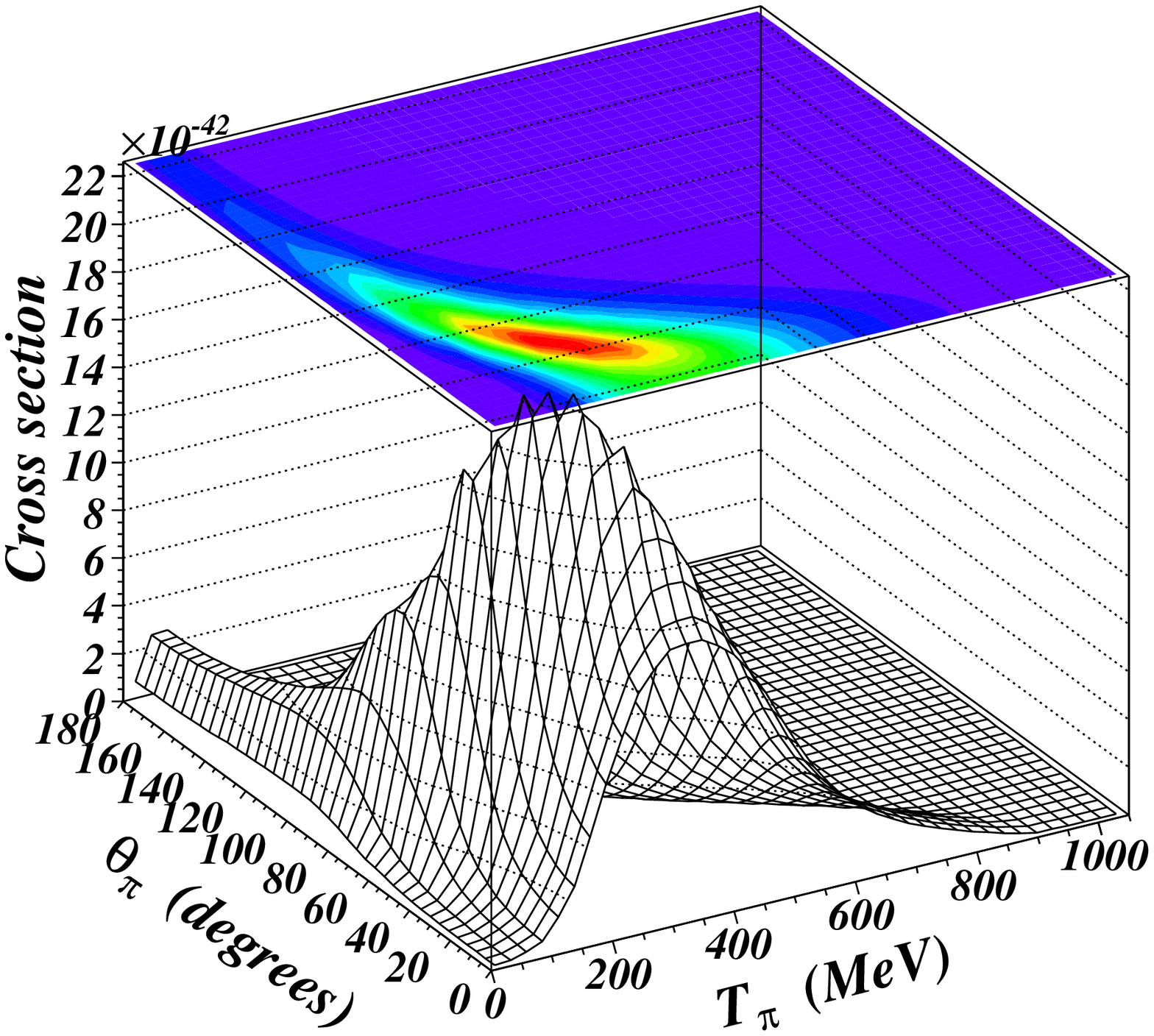}
  \caption{ Two-fold distributions for the process $\nu_{\mu} + p
  \rightarrow \mu + \Delta^{++}$ on a carbon nucleus for $E_{\nu} = 1200$\
  MeV.}
\end{figure}

Lately, new cross-section measurements presented by the MiniBooNE and
K2K collaborations have put the spotlights on few-GeV
neutrino-scattering physics.  As nuclei serve as neutrino detectors in
these experiments, there is a great deal of interest in modeling
neutrino-nucleus interactions in the region $W < 2\ \mbox{GeV}$, 
where the vast part of the strength is due to quasi-elastic events and
resonant one-pion production.  The need for a realistic description of
nuclear effects becomes even more evident in the light of future
neutrino-scattering experiments like Miner$\nu$a, who aim at a precise
study of various exclusive channels with the use of high-intensity
beams and improved particle identification.\\
In earlier work, neutrino-induced one-nucleon knockout calculations
have been performed within the relativistic multiple-scattering
Glauber approximation \cite{Martinez}.  Here, we proceed along the same lines to develop a 
framework for resonant one-pion production calculations.  The
presented formalism focuses on an intermediate $\Delta$ state, but can be straightforwardly extended 
to the second-resonance region.\\
\ \ \ \\
For a nucleus with mass number $A$, the process under consideration can be schematically represented as
\begin{equation}
\label{process}
\nu + A \stackrel{ \Delta }{ \rightarrow } l + N + \pi + (A-1),
\end{equation}
with $l$, $N$ and $\pi$ representing the outgoing charged lepton,
nucleon and pion respectively.  In the laboratory system, the
eightfold cross section for the process \eqref{process} is given by
\begin{equation}  
\label{cs}
  \frac{d^8\sigma}{dE_l d\Omega_l dE_{\pi} d\Omega_{\pi} d\Omega_{N}} = \frac{m_l |\vec{k}_l|}{(2\pi)^3} \frac{M_N M_{A-1} |\vec{k}_{\pi}|
    |\vec{k}_{N}|}{2 (2\pi)^5 |E_{A-1} + E_N + E_N \vec{k}_N \cdot (\vec{k}_{\pi} - \vec{q})/|\vec{k}_N|^2|} \overline{\sum}_{if} | M_{fi} |^2,  
\end{equation}
using self-explanatory notations for the outgoing particles' kinematics.\\
All information about the reaction dynamics is contained in the matrix element  
\begin{equation}
\label{me}
M_{fi} = i \frac{ G_F \cos\theta_c }{ \sqrt{2} } \overline{u}(k_N,s_N)
\Gamma^{\mu}_{\Delta\pi N}(k_{\pi},k_{\Delta})
S_{\Delta,\mu\nu}(k_{\Delta})
\Gamma^{\nu\rho}_{WN\Delta}(k_{\Delta},q) S_{W,\rho\sigma}(q)
J^{\sigma}_l u_{\alpha,m}(k_i),
\end{equation}     
where $G_F$ and $\theta_c$ stand for the Fermi constant and the
Cabibbo mixing angle.  In (\ref{me}), we adopted the impulse
approximation.  The hit nucleon is represented by the bound-state spinor $u_{\alpha,m}(k_i)$, calculated as the Fourier transform of the bound-state wave functions
\begin{equation}
\label{bswf1}
\Psi_{\alpha,m}(\vec{r}) = \left( \begin{array}{c} i
  \frac{G(r)}{r} \mathcal{Y}_{+\kappa,m}(\hat{\vec{r}})  \\ 
- \frac{F(r)}{r} \mathcal{Y}_{-\kappa,m}(\hat{\vec{r}})
\end{array} \right).
\end{equation}     
The radial wave functions in (\ref{bswf1}) are determined in the Hartree approximation to the $\sigma-\omega$ Walecka model \cite{Serot}.  Further, $J_l$ represents the weak lepton current and $S_{W}$ is the weak boson propagator.  To describe the $\Delta$-production vertex $\Gamma_{WN\Delta}$, we turn to the phenomenological form-factor parameterization discussed in \cite{Lalakulich}.  The adopted form factors are constrained by theoretical principles like CVC and PCAC and, in the case of the vector form factors, by available electron-scattering data.  For the $\Delta$ propagator we take the Rarita-Schwinger propagator for a spin-3/2 particle.  In this regard, medium modifications of the resonance are accounted for by implementing a shift to the mass and width of the $\Delta$.  We hereby use a density-dependent parameterization suggested in \cite{Oset}, and based on a calculation of the $\Delta$ self energy in the medium.  Finally, the decay of the $\Delta$ particle is described by the interaction $\Gamma_{\Delta\pi N}$, and $\overline{u}(k_N,s_N)$ represents the outgoing nucleon's spinor.\\
Next to binding effects and medium-modified $\Delta$ properties, the
final-state interactions (FSI) of the escaping nucleon and pion can
have a considerable effect on the calculated cross-section strength.  To
compute the influence of FSI, we adopt a relativistic
multiple-scattering Glauber approximation (RMSGA) \cite{Ryckebusch}.
Within this RMSGA model, one computes the attenuation of \textit{fast}
nucleons and pions due to elastic and mildly inelastic collisions with
the remaining \textit{spectator} nucleons when they travel through the
nucleus.  The Glauber approach allows to calculate the probability
that a high-energy nucleon/pion will escape from a finite nucleus
\cite{Ryckebusch2, Cosyn}, a quantity often referred to as the nuclear
transparency.  In Ref.~\cite{Martinez}, it was shown that plane-wave
$(\nu,\nu' N)$ cross sections corrected with this nuclear transparency
factor provide an excellent alternative for full, unfactorized
distorted-wave calculations, provided that inclusive cross sections
are considered.\\
\ \ \ \\
In short, we have presented a fully relativistic formalism for
neutrino one-pion production on nuclei in the resonance region.  
This framework opens up a wide range of possibilities: we can do calculations for
different nuclei and resonances.  Moreover, predictions can be made
for various observables, including not only $Q^2$ and $W$
distributions,  but also energy and angular distributions for the
outgoing lepton or hadrons (Fig.~\ref{Preliminary}).  As an accurate description of nuclear
effects will be of notable interest to future neutrino-scattering
experiments, we account for nuclear binding effects, medium-modified
resonance properties and FSI effects \cite{Praet}. 

\begin{theacknowledgments}
  The authors acknowledge financial support from the Fund for
  Scientific Research (FWO) Flanders.
\end{theacknowledgments}



\bibliographystyle{aipproc}   


\begin{thebibliography}{9}

\bibitem{Martinez}
M.~C. Mart\'{i}nez, P.~Lava, N.~Jachowicz, J.~Ryckebusch, K.~Vantournhout, and J.~M. Ud\'{i}as, \emph{Phys. Rev.} \textbf{C73}, 024607 (2006).

\bibitem{Serot}
B.~D. Serot,  and J.~D. Walecka, \emph{Adv. Nucl. Phys.} \textbf{16}, 1-327 (1986).

\bibitem{Lalakulich}
O.~Lalakulich,  and E.~A. Paschos, \emph{Phys. Rev.} \textbf{D71}, 074003 (2005).

\bibitem{Oset}
E.~Oset, and L.~L. Salcedo, \emph{Nucl. Phys.} \textbf{A468}, 631-652 (1987).

\bibitem{Ryckebusch}
J.~Ryckebusch, D.~Debruyne, P.~Lava, S.~Janssen, B.~Van Overmeire, and T.~Van Cauteren, \emph{Nucl. Phys.} \textbf{A728}, 226-250 (2003).

\bibitem{Ryckebusch2}
J.~Ryckebusch, W.~Cosyn, B.~Van Overmeire, and M.~C. Mart\'{i}nez, \emph{Eur. Phys. J.} \textbf{A31}, 585-587 (2007).

\bibitem{Cosyn}
W.~Cosyn, M.~C. Mart\'{i}nez, J.~Ryckebusch, and B.~Van Overmeire,
\emph{Phys. Rev.} \textbf{C74}, 062201 (R) (2006).

\bibitem{Praet} 
C.~Praet, O.~Lalakulich, N.~Jachowicz, and J.~Ryckebusch, \emph{in preparation}.

\end{thebibliography}



\end{document}